\documentclass[twocolumn,showpacs,preprintnumbers,amsmath]{revtex4}
\usepackage{amssymb}

\usepackage{graphicx}
\usepackage{dcolumn}
\usepackage{bm}
\usepackage{dsfont}
\usepackage{amsmath}
\usepackage{amsthm}
\usepackage{amscd}

\begin{document}

\title{
Constraints for weakly interacting light bosons \\from existence of massive neutron stars
}

\author{M.~I.~Krivoruchenko$^{1}$,~F.~\v Simkovic$^{2,3}$,~Amand~Faessler$^{4}$}
\affiliation{$^{1}$Institute for Theoretical and Experimental Physics$\mathrm{,}$ B. Cheremushkinskaya 25\\
117218 Moscow, Russia}
\affiliation{$^{2}$Bogoliubov Laboratory of Theoretical Physics$\mathrm{,}$ JINR \\
141980 Dubna$\mathrm{,}$ Moscow Region$\mathrm{,}$ Russia}
\affiliation{$^{3}$Department of Nuclear Physics and Biophysics$\mathrm{,}$ Comenius University$\mathrm{,}$ Mlynsk\'a dolina F1\\ SK--842 48 Bratislava$\mathrm{,}$ Slovakia}
\affiliation{$^{4}$Institut f\"{u}r Theoretische Physik$\mathrm{,}$ T\"{u}bingen Universit\"{a}t$\mathrm{,}$ Auf der Morgenstelle 14 \\
D-72076 T\"{u}bingen, Germany}

\begin{abstract}
Theories beyond the standard model include a number of new particles some of which might be light 
and weakly coupled to ordinary matter. Such particles affect the equation of state of nuclear matter 
and can shift admissible masses of neutron stars to higher values. The internal structure of neutron 
stars is modified provided the ratio between coupling strength and mass squared of a weakly interacting 
light boson is above $g^2/\mu^2 \sim 25 ~\mathrm{GeV}^{-2}$. We provide limits on the couplings with 
the strange sector, which cannot be achieved from laboratory experiments analysis. When the couplings 
to the first family of quarks is considered the limits imposed by the neutron stars
are not more stringent than the existing laboratory ones. The observations on neutron stars give evidence 
that equation of state of the $\beta$-equilibrated nuclear matter is stiffer than expected from 
many-body theory of nuclei and nuclear matter. A weakly interacting light vector boson coupled predominantly 
to the second family of the quarks can produce the required stiffening.
\end{abstract}

\maketitle   

Dark energy explains the accelerating expansion of the Universe.
The density of dark energy $\rho_D \approx 3.8$ keV/cm$^3$
may correspond to a fundamental scale $\lambda_D = \rho_D^{-1/4} \approx 8.5
\times 10^{-5}$ m \cite{BEAN97,Sundrum:1997js,ARKA98,DVAL01}. 
Theoretical schemes with extra dimensions suggest modifications of gravity 
below $\lambda_D$ and a multitude of states with masses above $ 1/\lambda_D$
very weakly coupled to members of multiplets of the standard model.
Scales significantly below $\lambda _{D}$ represent the interest for supersymmetric
extensions of the standard model which include generally a number of new
particles, such as the leading dark matter candidate neutralino. Typically,
new particles are expected with masses above several hundred GeVs or even
higher. However, light particles may exist also, such as a
neutral very weakly coupled spin-1 gauge $U$-boson \cite{FAYE80} that can provide
annihilation of light dark matter and be responsible
for the 511 keV line observed from the galactic bulge \cite{JEAN03,BOEH04}.

Deviations from the inverse-square Newton's law are parametrized often in
terms of the exchanges by hypothetical bosons also. Constraints on the
deviations from Newton's gravity have been set experimentally in the
sub-millimeter scale \cite{KAPN07,GERA08,LAMO97,DECC05,MOST08,BORD94} and
down to distances $\sim 10$ fm where effects of light bosons of extensions 
of the standard model can be expected \cite
{BARB75,POKO06,NESV04,NESV08,KAMY08}. Constraints on the coupling constants
from unobserved missing energy decay modes of ordinary mesons are discussed
in Ref. \cite{FAYE06}.

Bosons with small couplings escape detection in most
laboratory experiments. However, bosons interacting with
baryons modify the equation of state (EOS) of nuclear
matter. Their effect depends on the ratio between the
coupling strength and the boson mass squared, so a weakly
interacting light boson (WILB) may influence the structure
of neutron stars even if its baryon couplings are very small.

The effect of a vector boson on the energy density of nuclear matter can be
evaluated by averaging the corresponding Yukawa potential: 
\begin{equation}
E_{I}=\frac{1}{2}\int d\mathbf{x}_{1}d\mathbf{x}_{2}\rho (\mathbf{x}_{1})%
\frac{g^{2}}{4\pi }\frac{e^{-\mu r}}{r}\rho (\mathbf{x}_{2}),  \label{1}
\end{equation}
where $\rho (\mathbf{x}_{1})=\rho (\mathbf{x}_{1})\equiv \rho $ is the
number density of homogeneously distributed baryons, $r=|\mathbf{x}_{2}-%
\mathbf{x}_{1}|$, $g$ is the coupling constant with baryons, and $\mu $ is
the boson mass. A simple integration gives 
\begin{equation}
E_{I}=V\frac{g^{2}\rho ^{2}}{2\mu ^{2}},  \label{OMEGA}
\end{equation}
where $V$ is the normalization volume.

The coherent contribution to the energy density of nuclear matter from vector WILBs
should be compared to that from the ordinary $\omega$-mesons. 
In one-boson exchange potential  (OBEP) models,
the nucleon-nucleon repulsive core at short distances $r \lesssim b = 0.4$
fm is attributed to $\omega$-meson exchanges. Respectively, the $\omega$%
-meson plays a fundamental role in nuclear matter EOS. In the mean-field
approximation, the contribution of $\omega$-meson exchanges to the energy
has the form of Eq.(\ref{OMEGA}), with $g$ and $\mu$ replaced by the $\omega$%
-meson coupling $g_{\omega}$ and the mass $\mu_{\omega}$.

The $NN$ interactions are described with ${g_{\omega}^2}/{\mu_{\omega}^2} =
175 ~\mathrm{GeV}^{-2}$ \cite{STOC99}. The relativistic mean field (RMF)
model \cite{CHIN74} gives ${g_{\omega}^2}/{\mu_{\omega}^2} = 196 ~\mathrm{GeV%
}^{-2}$. The compression modulus of nuclear matter $K=210\div 300$ MeV is
consistent with ${g_{\omega}^2}/{\mu_{\omega}^2} = 125 \div 180 ~\mathrm{GeV}%
^{-2}$ \cite{GLEN96}. Stiff RMF models use ${g_{\omega}^2}/{\mu_{\omega}^2}$
up to $300 ~\mathrm{GeV}^{-2}$ \cite{TUCH05}. If we wish to stay within
current limits and do not want to modify the internal structure of neutron
stars qualitatively, as described by realistic models of nuclear matter, one has to
require that vector WILBs fulfill constraint 
\begin{equation}
\frac{g^2}{\mu^2} \lesssim \frac{g_{\omega}^2}{\mu_{\omega}^2} \approx 200 ~%
\mathrm{GeV}^{-2}.  \label{vLIMIT}
\end{equation}

A similar reasoning applies to scalar WILBs which have to compete with the
standard $\sigma$-meson exchange. In OBEP models, the long-range
attraction between nucleons is attributed to $\sigma$-meson exchanges. The
contribution of the $\sigma$-mesons to the interaction energy has the form
of Eq.(\ref{OMEGA}), with $g$ and $\mu$ replaced by the $\sigma$-meson
coupling $g_{\sigma}$ and the mass $\mu_{\sigma}$. The sign of the
contribution must be negative because of the attraction. Also, $\rho$ should be
replaced by the scalar density. In RMF models, the $\sigma$-meson mean field
decreases the nucleon mass. The effect depends on the ratio $g^2 /\mu^2$
also and produces an additional decrease of the energy at fixed volume and
baryon number. The empirical values of the ratio ${g_{\sigma}^2}/{%
\mu_{\sigma}^2}$ are $40 \div 60 \%$ higher than those of the $\omega$-meson 
\cite{STOC99,CHIN74,GLEN96,TUCH05}. The internal structure of neutron stars
is not modified significantly provided the coupling strength $g$ and mass $\mu$ of scalar
WILBs fulfill constraint 
\begin{equation}
\frac{g^2}{\mu^2} \lesssim \frac{g_{\sigma}^2}{\mu_{\sigma}^2} \approx 300 ~%
\mathrm{GeV}^{-2}.  \label{sLIMIT}
\end{equation}

The deviations from the Newton's gravitational potential are usually
parametrized in the form 
\begin{equation}
V(r) = - \frac{G m_1 m_2 }{r}\left(1 + \alpha_G e^{-r/\lambda} \right).
\label{NEW}
\end{equation}
The second Yukawa term can be attributed to new bosons with $G m^2 \alpha_G
= \pm g^2/(4 \pi)$ and $\lambda = 1/\mu$, 
where $+/-$ stands for scalar/vector bosons and $m$ is the proton mass.

On Fig. 1 we show regions in the parameter spaces $(g^2,\mu)$ and $%
(\alpha_G,\lambda)$ allowed for WILBs by the constraint (\ref{vLIMIT}). The
constraint for scalar bosons is close to (\ref{vLIMIT}). Constraints from
other works \cite
{BORD94,LAMO97,BARB75,NESV04,DECC05,MOST08,NESV08,POKO06,KAMY08} are shown
also.

\begin{figure}[!htb]
\begin{center}
\includegraphics[angle = 0,width = 7 cm]{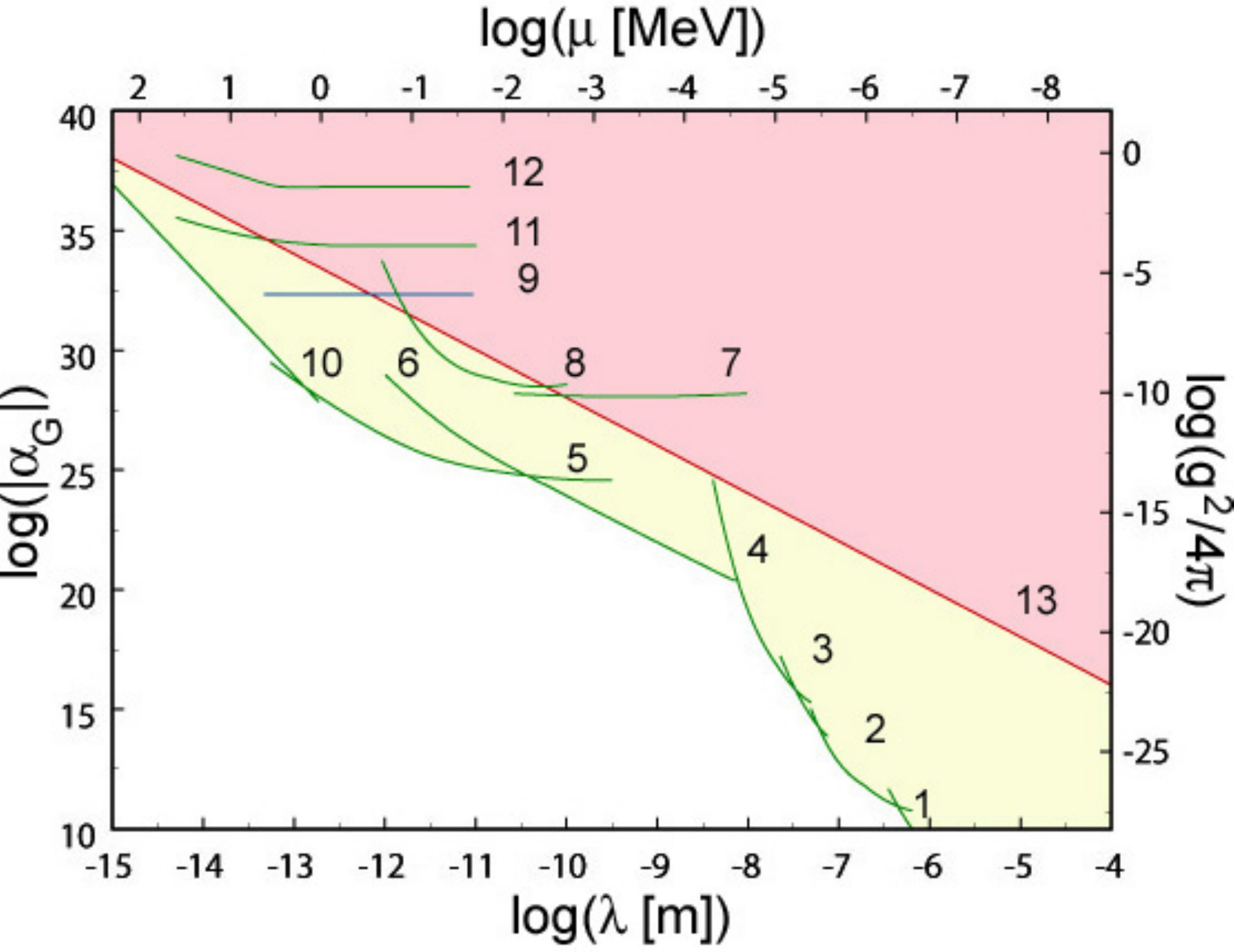}
\end{center}
\caption{(color online) Constraints on the coupling strength with nucleons $%
g^2/(4\pi)$ and the mass $\mu$ (equivalently $\alpha_G$ and $\lambda$) of
hypothetical weakly interacting light bosons: \textit{1} are constraints
from Ref. \protect\cite{LAMO97}, \textit{2} - from Ref. \protect\cite{DECC05}%
, \textit{3} - from Ref. \protect\cite{MOST08}, \textit{4} - from Ref. 
\protect\cite{BORD94}, \textit{5} and \textit{10} are constraints from
low-energy $n-^{208}$Pb scattering \protect\cite{POKO06} and \protect\cite
{BARB75}, respectively, \textit{6} - from Ref. \protect\cite{NESV08}, 
\textit{7} - from Ref. \protect\cite{NESV04}, \textit{8} and \textit{9} are
constraints from spectroscopy of antiproton atoms \protect\cite{POKO06}, 
\textit{11} and \textit{12} are constraints from near-forward $pn$
scattering for vector and scalar bosons, respectively \protect\cite{KAMY08}.
The axes are in the $\log_{10}$ scale. The internal structure of neutron
stars is not modified qualitatively provided the boson coupling strengths 
with baryons and masses lie at ${g^2}/{\mu^2} < 200 ~\mathrm{GeV}^{-2}$ beneath the
highlighted area \textit{13}. }
\end{figure}


An increase of $g$ (a decrease of $\mu$) of scalar WILBs increases the
negative contribution to pressure, makes EOS of nuclear matter softer, makes
neutron stars less stable against gravitational compression. The ratio $%
g^2/\mu^2$ cannot be increased significantly above the limit (\ref{sLIMIT}),
since the maximum mass of the neutron star sequence cannot be moved below
masses of the observed pulsars.

An increase of $g$ (a decrease of $\mu$) of vector WILBs, conversely,
increases the positive contribution to pressure, makes EOS of nuclear matter
stiffer, makes neutron stars more stable against gravitational compression
and drives the maximum mass of neutron stars up.

In case of vector bosons, it is less obvious what kind of the observables
confronts to high ratios $g^2/\mu^2$.


Realistic models of nuclear matter are based on the nucleon-nucleon
scattering data. They split into soft and stiff models according to the rate
the pressure increases with the density. The soft models correspond to low
maximum masses of neutron stars $\sim 1.6~\mathrm{M_{\odot}}$, while the
stiff models give the upper limit around $\sim 2.6~\mathrm{M_{\odot}}$.

The problem on the softness of nuclear EOS has received new interest due the
analysis of strange particle production in heavy-ion collisions. The data at
different bombarding energies lead to the conclusion that EOS of nuclear
matter must be soft at densities two to three times of the saturation
density \cite{FUCH01,HART06,FUCH07}. Data on the transverse and elliptic
flows in heavy-ion collisions suggest a soft EOS around the saturation, too 
\cite{DANI02}.

Last years observations of pulsars with high masses have been reported. The
most massive pulsars are PSR B1516+02B in the globular cluster M5 with the
mass of $1.96^{+0.09}_{-0.12}~\mathrm{M_{\odot}}$ and PSR J1748-2021B in the
globular cluster NGC 6440 with the mass of $2.74 \pm 0.22~%
\mathrm{M_{\odot}}$ \cite{FREI07}. The mass of rapidly rotating neutron star
in the low mass X-ray binary 4U 1636-536 is estimated to be $M=2.0 \pm 0.1 ~%
\mathrm{M_{\odot}}$ \cite{BARR05}. The mass and radius of the X-ray source
EXO 0748-676 are constrained to $M \geq 2.10 \pm 0.28~\mathrm{M_{\odot}}$
and $R \geq 13.8 \pm 1.8$ km \cite{OZEL06}. The observations on neutron
stars suggest that EOS of the $\beta$-equilibrated nuclear matter is stiff. 

The controversy between the conclusions on the softness of nuclear matter
as derived from the laboratory experiments and on the stiffness of
the $\beta$-equilibrated nuclear matter as derived from the astrophysical observations has been
of interest since after the discovery of millisecond pulsars \cite
{FRIE88,BROW88} and earlier \cite{SHAP83}.

Current models use to match EOS of neutron matter with a soft EOS at the
saturation density and a stiff EOS at higher densities. Such models are in the
qualitative agreement with laboratory and astrophysical data \cite{KLAN06}.

High densities provide favorable conditions for the occurrence of exotic
forms of nuclear matter: pion, kaon, and dibaryon condensates, quark matter.
New degrees of freedom make EOS softer, pushing the maximum mass of neutron
stars down. The recent astrophysical observations seem to exclude the softest EOS
e.g. based on the classical Reid soft core model \cite{PAND71} and make it
problematic to accommodate the exotic forms of nuclear matter with masses and
radii of the observed pulsars \cite{OZEL06} (see however \cite{ALFO06}).

The in-medium masses of vector mesons depend on the density. Assuming $\mu $
is a function of $\rho $ and using Eq.(\ref{OMEGA}), one may evaluate the $%
\omega $-meson contribution to pressure: 
\begin{equation}
P_{I}=\frac{g^{2}\rho ^{2}}{2\mu ^{2}}\left( 1-\frac{2\rho }{\mu }\frac{%
\partial \mu }{\partial \rho }\right) .  \label{PRESS}
\end{equation}
A positive shift of the $\omega $-meson mass decreases the pressure and
leads to a softer EOS, whereas a negative shift leads to a stiffer EOS. The
data on the dilepton production in heavy-ion collisions do not give evidence
for significant mass shift \cite{SANT08}, so the observed stiffness of
the $\beta$-equilibrated nuclear matter can hardly be attributed to in-medium modifications
of the vector mesons.

The realistic models of neutron matter discussed in Ref. \cite{KLAN06}
neglect hyperon channels e.g. reactions $\Sigma^- \to n + e + \bar{\nu}_e$.
In RMF models \cite{GLEN91,GLEN96,BURG03}, the $\beta$-equilibrium of
hyperons drops the limiting mass by $0.5 \div 0.8~\mathrm{M_{\odot}}$. This
result is in accord with hypernuclear data and other recent calculations 
\cite{ISHI08,SCHA08,DAPO08}. The inclusion of the $\beta$-equilibrium for
all baryons brings difficulties in reproducing the observed masses of neutron stars. 

Coming back to vector WILBs, we see that their existence is desirable to
provide additional stiffening of the $\beta$-equilibrated nuclear matter. 

The Compton wavelength of WILBs is assumed to be greater that the radius of nuclei
e.g. $1/\mu > R \approx 7$ fm $\approx (30 \;\mathrm{MeV})^{-1}$ for the lead. 
The contribution of WILBs to the binding energy of nuclei then equals
$\sim A^2 g^2/R$ like for photons. Since $g^2/(4\pi)$ is much smaller than the 
fine structure constant, the effect of WILBs on nuclei is negligible. 
Above $\sim 10^2$ MeV the coupling constant of WILBs is close to
unity, so WILBs there are neither weekly interacting nor light. 

WILBs thus do not modify observables in laboratory experiments on hypernuclear physics, nuclear structure
and heavy-ion collisions, since their baryon couplings are very small. The
characteristic scale of the parameters of these particles is fixed by
the upper limit (\ref{vLIMIT}).

\begin{figure}[!htb]
\begin{center}
\includegraphics[angle = 0,width = 7 cm]{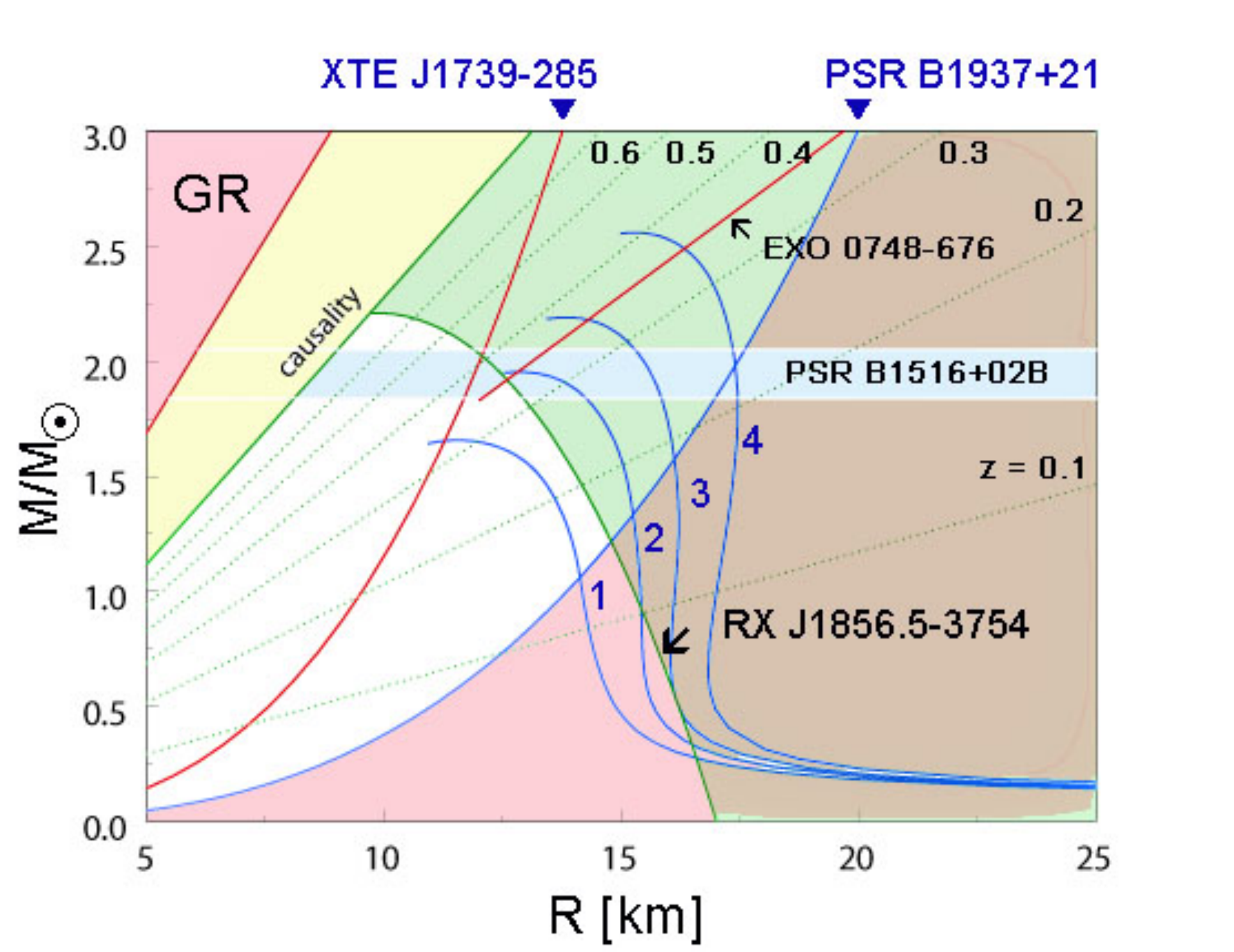}
\end{center}
\caption{(color online) Mass of non-rotating neutron stars as a function of
radius: \textit{1} - RMF model of hyperon mater with the compression modulus 
$K=300$ MeV \protect\cite{GLEN96}; \textit{2} - the same as \textit{1}
including a flavor-singlet vector WILB coupled to baryons with $g^2/\mu^2 =
25~\mathrm{GeV}^{-2}$ ($1/8$ of the limit (\ref{vLIMIT})); \textit{3} - the
same as \textit{2} with $g^2/\mu^2 = 50~\mathrm{GeV}^{-2}$; \textit{4} - the
same as \textit{2} with $g^2/\mu^2 = 100~\mathrm{GeV}^{-2}$. The highlighted
area within $M = 1.96^{+0.09}_{-0.12}~\mathrm{M_{\odot}}$ shows the mass
constraint from PSR B1516+02B. The neutron star sequences should cross the
rotation speed limit curves shown for pulsar PSR B1937+21 with the rotation
frequency of $\nu = 642$ Hz \protect\cite{ASHW83} and the neutron star XTE
J1739-285 showing X-ray burst oscillations with frequency of $\nu = 1122$ Hz 
\protect\cite{KAAR07}. The mass-dependent lower bound on radii of neutron
stars determined from the blackbody radiation of RX J1856.5-3754 is shown.
The dotted straight lines $z=0.1\div0.6$ indicate the red shift at surfaces
of neutron stars. The red shift of $z=0.35$ measured for EXO 0748-676
constrains the radii of neutron stars by $R > 12$ km and, respectively,
masses \protect\cite{OZEL06}. }
\end{figure}

The mass-radius relations for non-rotating neutron stars are shown on Fig. 2 for four
values of the ratio $g^2/\mu^2 = 0,~25,~50$ and $100~\mathrm{GeV}^{-2}$ of a
flavor-singlet vector WILB. At densities
below $\rho_{drip} = 4.3\times 10^{11} ~\mathrm{g/cm}^3$ the matter
represents an atomic lattice. WILBs do not modify properties of nuclei and
the Baym-Pethick-Sutherland EOS \cite{BPS}, accordingly. At densities $\rho_{drip} < \rho
\lesssim \rho_{nucl} = 2.8\times 10^{14} ~\mathrm{g/cm}^3$, atomic lattice
coexists with neutron liquid. The matter at $\rho_{drip} < \rho \lesssim
\rho_{nucl}$ is described by the Baym-Bethe-Pethick EOS \cite{BBP}. Above $\rho_{nucl}$,
nuclei dissolve and the matter is described by the $\beta$-equilibrated
hyperon liquid with the compression modulus $K=300$ MeV \cite{GLEN96}. WILBs
contribute to the energy density and pressure above $\rho_{drip}$, as
described by Eqs.(\ref{OMEGA}) and (\ref{PRESS}) with $\partial \mu/\partial
\rho = 0$, through the spatially extended nucleon and hyperon liquid
components of the neutron star matter. The vector WILBs give equal
contributions to the chemical potentials of the octet baryons and do not violate
the chemical $\beta$-equilibrium. 
\footnote{
The chemical potentials of leptons equal to the difference of the chemical
potentials of two baryons. For instance, the $\Sigma^- \to n + e + \bar{\nu}%
_e$ decay channel leads to the condition $\mu_e = \mu_{\Sigma^-} - \mu_{n}$.
The contribution of a flavor-singlet vector boson to the chemical potentials
of baryons cancels from $\mu_e$.
} 
The inclusion of such vector bosons does
therefore not change composition of the neutron star matter.

The highlighted area at the upper left corner of Fig. 2 excludes within
general relativity the radii of neutron stars below the Schwarzschild
radius. The causal limit excludes the area $R \lesssim 3GM$ \cite{LATT90}.
The rotation speed limit curves are constructed using the modified Keplerian
rate $\nu_{\max} \simeq 1045~ (M/\mathrm{M_{\odot}})^{1/2} (10 ~\mathrm{km}%
/R)^{3/2}$ Hz, which accounts for the deformation of rotating neutron stars 
and effects of general relativity \cite{LATT04}.

It is seen from Fig. 2 that, despite we selected EOS with the high
compression modulus, the neutron star sequence with $g^{2}/\mu ^{2}=0$
contradicts to the mass measurement of PSR B1516+02B. It gives a very low
mass of the neutron star from the blackbody radiation radius constraint 
also, which confronts with the lower limit of $\sim 0.85~\mathrm{M_{\odot}}$
for masses of protoneutron stars \cite{GOND}.

The value of $g^2/\mu^2 = 200~\mathrm{GeV}^{-2}$ gives the maximum mass
slightly above $3.0~ \mathrm{M_{\odot}}$. However, the neutron star sequence
does not cross the rotation speed limits, while the red shift remains always
below $z=0.35$. The upper bound (\ref{vLIMIT}) is thus critical for the
internal structure of neutron stars. 
\footnote{
We do not discuss the gravitational mass - baryon mass relationship for PSR
J0737-3039 (B) \cite{PODS05}, since majority of realistic models fail to
reproduce it. None of the neutron star models with WILBs fits the radiation
radius $R_{\infty}=12.8 \pm 0.4$ km of an X-ray source in the globular
cluster M13 \cite{GEND03}. Also, the rotation speed limit from the X-ray
transient XTE J1739-285 that favors a soft EOS and the mass of pulsar PSR
J1748-2021B that favors a very stiff EOS are nearly mutually exclusive.
These data need confirmation.}

The vector WILBs increase the minimum and maximum mass limits and
radii of neutron stars and are able to bring in the agreement models of
hyperon matter which are soft with the astrophysical observations on neutron
stars which require a stiff EOS. The ratio $g^2/\mu^2 \approx 50~\mathrm{GeV}%
^{-2}$ might be reasonable. Such a value, however, clearly
contradicts to the laboratory constraints shown on Fig. 1 in the entire mass
range $\mu = 10^{-9}$ to $10^{2}$ MeV. 

The in-medium modification of masses of vector bosons modify EOS. Vector
WILBs can be compared to the $\omega$-meson where $|\delta
\mu_{\omega}|/\mu_{\omega} \lesssim 0.1$ above the saturation density \cite
{SANT08}. A vector WILB mass shift can be estimated as $\delta \mu^2 \sim
g^2/g^2_{\omega} 2 \mu_{\omega} \delta \mu_{\omega}$. The in-medium
modification is small provided $|\delta \mu^2| \lesssim \mu^2$ i.e. $%
g^2/\mu^2 \lesssim 10^3 ~\mathrm{GeV}^{-2}$, so in the region of interest 
(\ref{vLIMIT}) holds for the vacuum masses.

The laboratory constraints shown on Fig. 1 do not apply to WILBs coupled to 
hyperons. A vector WILB coupled predominantly to the second family of 
the quarks makes hyperon matter EOS stiffer also. It contributes differently 
to chemical potentials of the octet baryons and suppresses the hyperon content 
of the neutron star matter due the additional repulsion. One can expect the ratio 
$g^{2}/\mu ^{2}$ should be close to or higher than that estimated above 
($\sim 50~\mathrm{GeV}^{-2}$). In such a scenario, nuclear matter without hyperons 
can be treated as reasonable approximation for the modeling structure of neutron stars in the 
$\beta $-equilibrium also e.g. on line with Ref. \cite{KLAN06} where models
with the blocked hyperon channels are shown to be in the qualitative agreement with the
laboratory and astrophysical constraints. 

Gauge bosons interact with the conserved currents only,
but flavor is not conserved. A WILB coupled to the second
family of the quarks cannot be a gauge boson, so it does not
arise naturally in the current theoretical schemes. Here, we
do not have a goal whatsoever to go beyond the phenomenological
analysis.

Hypernuclear data restrict $NY$ potentials, whereas the
interaction between hyperons $ YY$ is not known experimentally.
The stiffness of the hyperon matter might also be
attributed to the $\phi(1020)$-meson exchange, whose coupling
to the nonstrange baryons is suppressed according
to the Okubo-Zweig-Iizuka rule (see, however, \cite{Genz:1976fk}).

Summarizing, we have assumed the existence and derived
constraints for a new boson that couples to nuclear
matter. Such a particle contributes, by its coherent force
among nuclear constituents, to a modified EOS and affects
the structure of neutron stars. The neutron stars exclude
scalar bosons with the coupling strengths and masses
above the line \textit{13} on Fig. 1, whereas in a narrow band
below it and above a vector boson coupled to quarks of the
second family could modify the EOS in a direction favored
by the observed masses and radii of neutron stars. The
astrophysical constraints in the nonstrange sector are less
stringent than the most accurate laboratory ones. They are
unique, however, for scalar WILBs in the strange sector.
The region of validity of the astrophysical constraints
extends from $\lambda \sim 10$ fm to about 10 km. Detailed studies
of manifestations of new bosons in astrophysics, physics of
neutron stars, and hadron decays to energy missing channels
can shed more light on the existence of WILBs and
their possible effect on the structure of neutron stars.

\begin{acknowledgments}

The authors are grateful to S.~G.~Kovalenko and Yu.~N.~Pokotilovski for useful discussions. 
This work is supported 
by the EU ILIAS project under the contract RII3-CT-2004-506222, 
the VEGA Grant agency under the contract  No.~1/0249/03, DFG grant 
No. 436 RUS 113/721/0-3, and RFBR grant No. 09-02-91341.

\end{acknowledgments}

\end{document}